\documentstyle[conf_iap,epsf]{article}

\def\avrg#1{{\langle #1 \rangle}}

\def\hmpc{{\rm\, h^{-1}Mpc}}

\def\kpc{{\rm\,kpc}}
\def\mpc{{\rm\,Mpc}}
\def\msun{{\rm\,M_\odot}}
\def\eg{{\it e.g., }}
\def\ie{{\it i.e., }}
\def\etal{{\it et al. }}
\def\etc{{\it etc. }}

\def\spose#1{\hbox to 0pt{#1\hss}}
\def\lta{\mathrel{\spose{\lower 3pt\hbox{$\mathchar"218$}}
     \raise 2.0pt\hbox{$\mathchar"13C$}}}
\def\gta{\mathrel{\spose{\lower 3pt\hbox{$\mathchar"218$}}
     \raise 2.0pt\hbox{$\mathchar"13E$}}}

\begin{document}

\heading{Theoretical Tools for Large Scale Structure}
\par\medskip\noindent

\author{J. Richard Bond$^{1}$, Lev Kofman$^{1}$, Dmitry Pogosyan$^{1}$ and James Wadsley$^{1,2}$}
\address{
Canadian Institute for Theoretical Astrophysics, University of Toronto,\\
\mbox{\hspace{3mm}}60 St. George St., Toronto, ON M5S 3H8, Canada}
\address{Astronomy Department, Univ. of Washington,  Box 351580,
Seattle WA 98195-1580}

\begin{abstract}
We review the main theoretical aspects of the structure formation
paradigm which impinge upon wide angle surveys: the early universe
generation of gravitational metric fluctuations from quantum noise in
scalar inflaton fields; the well understood and computed linear regime
of CMB anisotropy and large scale structure (LSS) generation; the
weakly nonlinear regime, where higher order perturbation theory works
well, and where the cosmic web picture operates, describing an
interconnected LSS of clusters bridged by filaments, with membranes as
the intrafilament webbing. Current CMB+LSS data favour the simplest
inflation-based $\Lambda$CDM models, with a primordial spectral index
within about 5\% of scale invariant and $\Omega_\Lambda \approx 2/3$,
similar to that inferred from SNIa observations, and with open CDM
models strongly disfavoured. The attack on the nonlinear regime with a
variety of N-body and gas codes is described, as are the excursion set
and peak-patch semianalytic approaches to object collapse. The
ingredients are mixed together in an illustrative gasdynamical
simulation of dense supercluster formation.
\end{abstract}

\section{Introduction}

\begin{figure}[!ht]
\epsfxsize=5.0in\leavevmode\epsffile{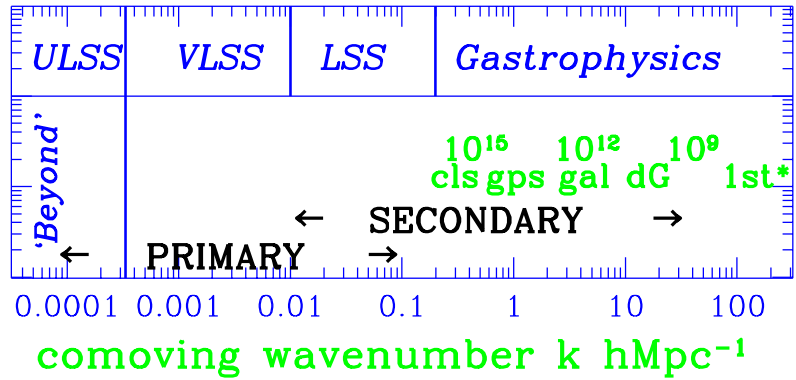}
\vspace{-0.8in}
\caption{This shows the bands in comoving wavenumber $k$ probed by CMB
primary and secondary anisotropy experiments and LSS observations. The
nonlinear wavenumber $k_{NL}(t)$, where the mass density power crosses
unity, propagates from high to low $k$ with redshift leaving in its
wake collapsed halos and at its bow the interconnected cosmic web
pattern replayed with time on progressively increasing scales. The
objects form from waves concentrated in the $k$-space bands that their
labels cover: the first star forming tiny dwarf galaxies appear (``1st
*''), typically at redshift about 20; dwarf galaxies (dG); normal
galaxies (gal); groups (gps); and clusters (cls), where we are
now. Equivalent mass scales are given above them. Secondary
anisotropies arise only once matter has gone nonlinear.  The range
extends to the LSS regime because of clustering of the nonlinear
objects, \eg of clusters for the SZ effect. }
\label{fig:probes1} 
\end{figure}

By their very nature, wide angle surveys, whether of CMB anisotropies
or galaxy redshifts, are designed to probe long wavelengths in the
universe, \ie low comoving Fourier wavenumbers $k$
(Fig.~\ref{fig:probes1}). We are hoping for simplicity of
interpretation, best if the observables are probes of linear physics,
harder to disentangle if the probe is of the dissipative chaotic
physics characterizing the highly nonlinear regime, and hopefully
tractable in the mildly nonlinear regime. The goal is to extract the
underlying linear gravitational metric fluctuation field $\Phi$ which
characterizes the initial conditions and the cosmic parameters that
define its evolution, whatever the specific CMB or density-based
observable.

A theorist would argue that it is best to go for CMB probes, using
spectacular forecasts \cite{bet97,zss97} of the percent level accuracy
that cosmic parameters can be determined to because of the sensitive
dependence of the anisotropies on the linear gas physics at photon
decoupling at $z \sim 1000$. However, there are significant
near-degeneracies among cosmological parameters such as $\Lambda$ and
the curvature, which require other experiments, often wide-angle, to
break. Further, to realize the promise will depend upon how well
foreground signals from the Milky Way and secondary anisotropy signals
from nonlinear phenomena such as the Sunyaev-Zeldovich effect from
high pressure gas and emission from dusty high redshift galaxies can
be disentangled from our goal, the primary anisotropies. CMB surveys
can deliver useful information all the way from the ultralarge scale
structure (ULSS) realm ``beyond'' (our Hubble patch) through the
highly linear VLSS regime well-probed by COBE, into the LSS regime. We
are now getting increasingly good LSS information from current small
angle CMB anisotropy experiments (bandpower estimates shown in
Fig.~\ref{fig:CLth}), we are about to take a significant step forward
with balloon-borne and interferometer experiments, and a leap forward
with the satellites MAP and Planck (lower right panel of
Fig.~\ref{fig:CLth}).

\begin{figure}
\vspace{-1.0in}
\begin{center}
\epsfysize=5.0in\leavevmode\epsffile{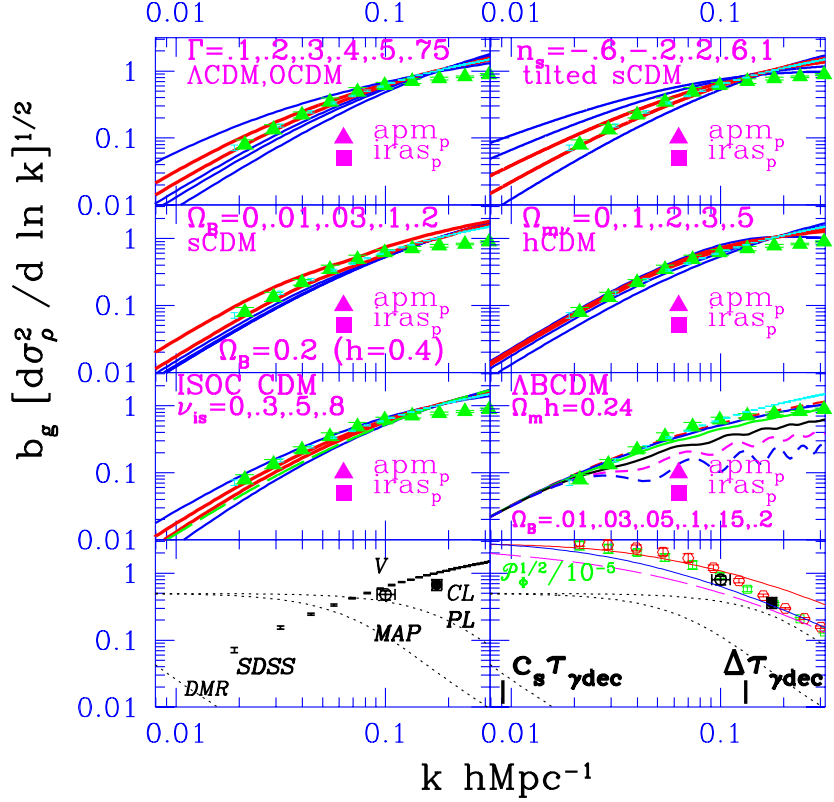}
\vspace{-0.4in}
\end{center}
\caption{These panels show how the shape of the linear density power
spectrum in the LSS regime changes as the cosmological model is
varied. These are compared with the reconstructed linear power
spectrum estimated by Peacock\cite{peacock96}. It is unclear how
seriously the discrepencies above $0.1 (\hmpc )^{-1}$ should be
taken. The two lower panels show how strong the overlap will be
between CMB and redshift surveys.  Right focuses on ${\cal
P}_\rho^{1/2}$, left on ${\cal P}_\Phi^{1/2}$. A forecast of error
bars for the number density power spectrum derived from the SDSS by
Vogeley and Szalay, and current mass density power spectrum estimates
from cluster abundances ($CL$) and large scale streaming velocities
($V$), are contrasted with the dotted $k$-space filters for MAP and
Planck. Planck in particular encompasses well the two important scales
which define the effective acoustic peak range for primary anistropies
(those involving linear fluctuations): the sound crossing distance
$c_s\tau_{\gamma dec}$ at photon decoupling around redshift 1000; the
width $\Delta \tau_{\gamma dec}$ of the region over which this
decoupling occurs, which is about a factor of 10 smaller, and below
which primary CMB anisotropies are damped.  Estimates of the linear
$\Phi$ power from current galaxy clustering data by
{\protect\cite{peacock96}} are compared with some sample (linear)
COBE-normalized gravitational potential power spectra on the right.  A
bias is ``allowed'' to (uniformly) raise the shapes to match the
observations.  }
\label{fig:smprobes} 
\end{figure}

Galaxy redshift surveys probe number density fields, so they are
largely limited to probing the LSS band since the level of mass
density fluctuations is apparently rather tiny in the VLSS band. The
further complication inherent in biasing relative to the mass density
field may still obscure the derivation of the gravitational potential
field. In Fig.~\ref{fig:smprobes}, this unknown is absorbed in a
uniform-amplifier bias factor $b_g$ which the simplest biasing
theories predict in the linear regime, but which we know could be
quite environment and scale dependent in a way which may make reliable
calculations difficult.

We do think that for clusters the biasing has a simple physical
explanation involving collapsed halos that allows for reliable
calculations and extraction of the gravitational potential field from
cluster surveys constructed using X-ray, optical or SZ data. It is
this same simplicity of interpretation which has allowed the estimates
of the mass power spectrum from cluster abundances (lower panels of
Fig.~\ref{fig:smprobes}).

It seems that practical limitations will limit weak gravitational
lensing and large scale streaming velocities to the LSS regime, even
though they are direct probes of $\Phi$ gradients.

The length scales below LSS, labelled ``gastrophysics'', are ones over
which energy injection and propagation, through ionization and shock
fronts from galaxies, quasars, \etc are expected to have had important
effects. That is, even if the dark matter density power in the
$k$-bands in question (\eg the galaxy band of Fig.~\ref{fig:probes1}
at $z\sim 4$) is linear, the gas effects can lead to severe and
complex biasing or antibiasing over such scales. 
Thus, although probes
based on wide field catalogues of high redshift galaxies and quasars,
and on quasar absorption lines from the intergalactic medium,
represent a very exciting observational frontier, it will be difficult
for theoretical conclusions about the early universe and the
underlying fluctuations to be divorced from these gastrophysical
complications.

\subsection{Cosmic Structure and the Nonlinear Wavenumber} \label{seckNL}

Various (linear) density power spectra, ${\cal P}_\rho (k) \propto
k^4{\cal P}_{\Phi}(k)$, are shown in Fig.~\ref{fig:smprobes}. Many
people plot ${\cal P}_\rho/k^3 \sim k{\cal P}_\Phi$ instead. Here $
{\cal P}_{\Phi}(k)$ $ \equiv d\sigma_\Phi^2 /d\ln k$ is the {\it rms}
power in each $d\ln k$ band. In hierarchical structure formation
models such as those considered here, the nonlinear wavenumber
$k_{NL}(t)$, defined by $\sigma_\rho (k_{NL})=1$, where
$\sigma_\rho^2(k) \equiv \int_0^{k} {\cal P}_\rho (k)d\ln k$, grows
as the universe expands. $k_{NL}(t)$ was in the galaxy band at
redshift 3 and is currently in the cluster band.

At a given time $t$, there is a band in $k$ extending just above
$k_{NL}(t)$ which is primarily responsible for the nonlinear collapsed
dark matter halos in the medium in hierarchical theories. Smaller
halos that formed earlier from much higher $k$ bands would have
largely merged into the halos of relevance at epoch $t$, in a sequence
of characteristic objects shown in Fig.~\ref{fig:probes1}. If we
denote the {\it rms} linear density fluctuation level for waves longer
than $k$ by $\sigma_\rho(k)$, then the characteristic scale $k_*$
which gives the peak of the Press-Schechter distribution of collapsed
objects, $d\Omega_{coll} (M)/d\ln M$, occurs at $\sigma_\rho (k_*)=
f_c$, where $f_c$ is a suitable collapse threshold, the famous 1.686
for $\Omega=1$. Associated with $k_*$ is a characteristic
mass\footnote{The relation between $k$ and $M=(4\pi/3)\bar{\rho}_0
R_{TH}^3$ is that the {\it rms} $\sigma_\rho$ is the same for
``top-hat filtering'' on scale $R_{TH}$ and ``sharp k-space
filtering'' on scale $k^{-1}$.} $M_*$.  On the other hand, for mass
scales below $\sigma_\rho = f_c/3 \approx 0.6$ not even $3\sigma $
peaks will have collapsed at this resolution. As we describe later,
this is the weakly nonlinear domain of the cosmic web of
interconnected structure: the view of the structure smoothed on these
scales reveals the characteristic patterns of filaments connecting
clusters, and membranes connecting filaments. Voids are rare density
minima which have opened up by gravitational dynamics and merged,
opposite to the equally abundant rare density maxima, the clusters, in
which the space collapses by factors of 5-10 and more.

At $k >k_{NL}(t)$, nonlinearities and complications associated with
dissipative gas processes can obscure the direct connection to the
early universe physics. Indeed, the reason galaxies are still around
today when $k_{NL}$ is cluster scale is that, although the outer halos
of the galaxies will have largely merged, gas cooling allows the
baryons and some of the dark matter to concentrate and survive as
independent beings in groups and clusters, breaking the hierarchy.

Most easily interpretable are observables probing the linear regime
now, $k < k_{NL}(t_0)$. CMB anisotropies arising from the linear
regime are termed primary; as Fig.~\ref{fig:probes1} shows, these
probe 3 decades in wavenumber, with the range defined by CMB damping
rather than by any nonlinear effects at $z\sim 1000$. LSS observations
at low redshift probe a smaller, but overlapping, range. We have hope
that $z\sim 3$ LSS observations, when $k_{NL}(t)$ was larger, can
extend the range, but gasdynamics can modify the relation between
observable and power spectrum in complex ways. Secondary anisotropies
of the CMB, those associated with nonlinear phenomena, also probe
smaller scales and the ``gastrophysical'' realm.

\subsection{Early Universe Connection and the Freedom in Inflation} \label{secinflation} 

The source of fluctuations to input into the cosmic structure
formation problem is likely to be found in early universe physics. The
contenders for generation mechanism are (1) ``zero point'' quantum
noise in scalar and tensor fields that must be there in the early
universe if quantum mechanics is applicable and (2) topological
defects which may arise in the inevitable phase transitions expected
in the early universe.

From CMB and LSS observations we hope to learn: the statistics of the
fluctuations, whether Gaussian or non-Gaussian; the mode, whether
adiabatic or isocurvature scalar perturbations, and whether there is a
significant component in gravitational wave tensor perturbations; the
power spectra for these modes, $ {\cal P}_{\Phi}(k), {\cal P}_{is}(k)
, {\cal P}_{GW}(k)$ as a function of comoving wavenumber $k$. As the
Universe evolves the initial shape of ${\cal P}_{\Phi}$ (nearly flat
or scale invariant) is modified by characteristic scales imprinted on
it that reflect the values of cosmological parameters such as the
energy densities of baryons, cold and hot dark matter, in the vacuum
(cosmological constant), and in curvature.  Often observables can be
expressed as weighted integrals over $k$ of the final state power
specta and thus can probe both $\Omega$'s and initial power amplitudes
and tilts. 

 Many variants of the basic inflation theme have been proposed,
sometimes with radically different consequences for ${\cal P}_\Phi (k)
\sim k^{1-n_s(k)}$, and thus for LSS, which is used in fact to highly
constrain the more baroque models.  The most likely inflation
possibilities are the simplest: (1) adiabatic curvature fluctuations
with nearly uniform scalar tilt over the observable range, slightly
more power to large scales ($0.8 \lta n_s \lta 1$) than ``scale
invariance'' ($n_s=1$) gives, a predictable nonzero gravity wave
contribution with tilt similar to the scalar one, and tiny mean
curvature ($\Omega_{tot}\approx 1$); (2) same as (1), but with a tiny
gravity wave contribution. 

An arguable rank-ordering of the more baroque inflation possibilities
(see \eg \cite{bh95,LLKCBA97} for references) is: (3) same as (1) but
with a subdominant isocurvature component of nearly scale invariant
tilt (the case in which isocurvature dominates is ruled out, but see
Figs.~\ref{fig:smprobes},\ref{fig:CLth}); (4) radically broken scale
invariance with weak to moderate features (ramps, mountains, valleys)
in the fluctuation spectrum (strong ones are largely ruled out); (5)
radical breaking with non-Gaussian features as well; (6) ``open''
inflation, with quantum tunneling or possibly a Hawking-Turok instanton
resulting in a negatively-curved (hyperbolic) space which inflates,
but not so much as to flatten the mean curvature; (7) quantum creation
of compact hyperbolic space from ``nothing''. It is quite debatable
which of the cases beyond (2) are more or less plausible, with some
claims that (4) is supersymmetry-inspired, others that (6) is not as
improbable as it sounds.

\subsection{Cosmic Parameters and the CMB} \label{seccosmicparam}

Even simple Gaussian inflation-generated fluctuations for structure
formation have a large number of early universe parameters we would
wish to determine: power spectrum amplitudes at some normalization
wavenumber $k_n$ for the modes present, $\{ {\cal P}_{\Phi}(k_n),
{\cal P}_{is}(k_n) , {\cal P}_{GW}(k_n) \}$; shape functions for the
``tilts'' $\{ \nu_s(k),\nu_{is}(k), \nu_t(k) \} $, usually chosen to
be constant or with a logarithmic correction, \eg $\nu_s(k_n),
d\nu_s(k_n)/d\ln k$. (The scalar tilt for adiabatic fluctuations,
$\nu_s (k) \equiv d\ln {\cal P}_{\Phi}/d \ln k$, is related to the
usual index, $n_s$, by $\nu_s = n_s-1$.)  The transport problem is
dependent upon physical processes, and hence on physical parameters. A
partial list includes the Hubble parameter ${\rm h}$, various mean
energy densities $\{ \Omega_{tot}, \Omega_B , \Omega_{\Lambda},
\Omega_{cdm }, \Omega_{hdm }\}{\rm h}^2$, and parameters
characterizing the ionization history of the Universe, \eg the Compton
optical depth $\tau_C$ from a reheating redshift $z_{reh}$ to the
present. Instead of $\Omega_{tot}$, it is becoming conventional to use
the curvature energy parameter, $\Omega_k\equiv 1-\Omega_{tot}$, thus
zero for the flat case. In this space, the Hubble parameter, ${\rm h}=
(\sum_j (\Omega_j{\rm h}^2 ))^{1/2}$, and the age of the Universe,
$t_0$, are functions of the $\Omega_j{\rm h}^2$. The density in
nonrelativistic (clustering) particles is
$\Omega_{nr}=\Omega_B+\Omega_{cdm}+\Omega_{hdm}$, denoted as well by
$\Omega_m$. The density in relativistic particles, $\Omega_{er}$,
includes photons, relativistic neutrinos and decaying particle
products, if any.  $\Omega_{er}$, the abundance of primordial helium,
\etc should also be considered as parameters to be determined. The
count is thus at least 17, and many more if we do not restrict the
shape of ${\cal P}_{\Phi}(k)$ through theoretical considerations of
what is ``likely'' in inflation models. For example, the ratio of
gravitational wave power to scalar adiabatic power is ${\cal
P}_{GW}/{\cal P}_{\Phi}\approx -(100/9)\nu_t/(1-\nu_t/2)$, with small
corrections depending upon $\nu_s-\nu_t$ \cite{bh95}.  If such a
relationship is assumed, the parameter count is lowered by one, and
other restrictions on most likely behaviour or prior probabilities
reflecting other observations or prejudices are used to reduce the set
further; \eg the forecasts for the MAP and Planck satellites
in \cite{bet97} used a 9 parameter set.

For a given model, the early universe ${\cal P}_{\Phi}$ is uniquely
related to late-time power spectrum measures of relevance for the CMB,
such as the quadrupole or averages over $\ell$-bands, and to LSS
measures, such as the {\it rms} density fluctuation level on the
$8\hmpc$ (cluster) scale, $\sigma_8$, so any of these can be used in
place of the primordial power amplitudes in the parameter set.

\begin{figure}
\vspace{-1.0in}
\begin{center}
\epsfysize=5.0in\leavevmode\epsffile{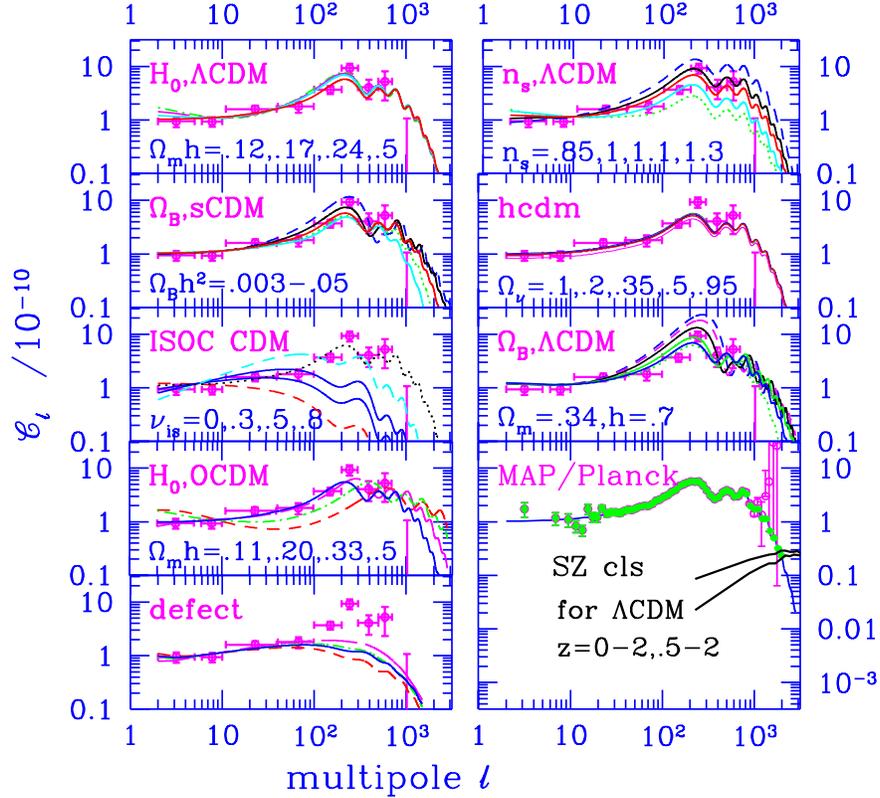}
\end{center}
\vspace{-0.4in}
\caption{The ${\cal C}_\ell$ anisotropy bandpower data up to summer
1998 compressed to 9 bands using the methods of {\protect\cite{bjk98}}
are compared with various 13 Gyr model sequences (left to right): (1)
$H_0$ from 50 to 90, $\Omega_{\Lambda}$, 0 to 0.87, for an untilted
$\Lambda$CDM sequence; (2) $n_s$ from 0.85 to 1.25 for the $H_0=70$
$\Lambda$CDM model ($\Omega_{\Lambda}=.66$) -- dotted is 0.85 with
gravity waves, next without, upper dashed is 1.25, showing visually why
$n_s$ is found to be unity to within 5\%; (3) $\Omega_B{\rm h}^2$ from
0.003 to 0.05 for the $H_0=50$ sCDM model; (4) $H_0$=50 sequence with
neutrino fractions varying from 0.1 to 0.95; (5) an isocurvature
CDM sequence with positive isocurvature tilts ranging from 0 to 0.8;
(6) $\Omega_B{\rm h}^2$ from 0.003 to 0.05 for the $H_0=70$
$\Lambda$CDM model; (7) $H_0$ from 50 to 65, $\Omega_{k}$ from 0 to
0.84 for the untilted oCDM sequence, showing the strong $\ell$-shift
of the acoustic peaks with $\Omega_k$; (9) sample defect ${\cal
C}_\ell$'s for textures, \etc from {\protect\cite{pst97}} -- cosmic
string ${\cal C}_\ell$'s from {\protect\cite{allen97}} are similar and
also do not fare well compared with the current data. The bottom right
panel is extended to low values to show the magnitude of secondary
fluctuations from the thermal SZ effect for the $\Lambda$CDM
model. The kinematic SZ ${\cal C}_\ell$ is significantly lower. Dusty
emission from early galaxies may lead to high signals, but the power
is concentrated at higher $\ell$, with a weak tail because galaxies
are correlated extending into the $\ell \lta 2000$ regime. Forecasts
of how accurate ${\cal C}_\ell$ will be determined for an sCDM model
from MAP (error bars growing above $\ell \sim 700$) and Planck (small
errors down the ${\cal C}_\ell$ damping tail) are also shown.}
\label{fig:CLth}
\end{figure}

The arena in which CMB theory battles observation is the anisotropy
power spectrum in multipole space, Fig.~\ref{fig:CLth}, which shows
how primary ${\cal C}_\ell$'s vary with some of these cosmic
parameters. Here ${\cal C}_\ell \equiv \ell (\ell +1) \avrg{\vert
(\Delta T/T)_{\ell m}\vert^2 }/(2\pi )$. Cosmological radiative
transfer in the linear regime and therefore the ${\cal C}_\ell$
predictions as the cosmological parameters are varied is on a firm
theoretical footing; for reviews, see
\cite{bh95,huSS97,huWhitepol97}. The ${\cal C}_\ell$ sets shown
correspond to those in Fig.~\ref{fig:smprobes}. They also are chosen
to have the fixed cosmological age of 13 Gyr, similar to that inferred
from globular cluster ages. Bond and Jaffe \cite{bj98} used all of the
current CMB data to test such sets for tilted $\Lambda$CDM,
h$\Lambda$CDM, oCDM sequences, with and without gravity wave
contributions and wide variation in $\Omega_\Lambda$, $H_0$, $n_s$,
\etc and for ages from 11 to 15 Gyr. Results quoted here on parameters
from current CMB data and combining CMB and LSS data are from that
work.

The ${\cal C}_\ell$'s are normalized to the 4-year {\it
dmr}(53+90+31)(A+B) data, which fixes it to within about 7\% in
amplitude.  DMR is fundamental to analyses of the VLSS region and ULSS
region, and is the data set that is the most robust at the current
time. Even with the much higher precision MAP and Planck experiments
we do not expect to improve the results on the COBE angular scales
greatly because the 4-year COBE data has sufficiently large
signal-to-noise that one is almost in the cosmic variance error limit
(due to realization to realization fluctuations of skies in the
theories) which cannot be improved upon no matter how low the
experimental noise.

The ``beyond'' land in Fig.~\ref{fig:probes1} is actually partly
accessible because ultralong waves contribute gentle gradients to our
CMB observables. The DMR data is well suited to testing whether
radical broken scale invariance results in a huge excess or deficit of power in
the COBE $k$-space band, {\it e.g.}, just beyond $k^{-1} \sim
H_0^{-1}$, but this has not been much explored. The remarkable
non-Gaussian structure predicted by stochastic inflation theory would
likely be too far beyond our horizon for its influence to be felt in
the DMR data. The bubble boundary in hyperbolic inflation models may
be closer and its influence shortly after quantum tunneling occurred
could possibly have observable consequences for the CMB. Theorists
have also constrained the scale of topology in simple models. For
compact spatial manifolds (which may have $\Omega \le 1 $ as well as
$\Omega >1 $), the wavenumbers have an initially discrete spectrum,
and are missing ultralong waves, limited by the size of the
manifold. Usually isotropy is radically broken in such models,
resulting in CMB pattern formation which allows much stronger
constraints to be imposed than results just from the lack of ULSS
power; \eg \cite{bpss98} find for flat equal-sided 3-tori, the
inscribed radius must exceed $1.1 (2H_0^{-1}) = 6600 \hmpc$ from DMR
at the $95\%$ confidence limit, with the weaker $> 0.7 (2 H_0^{-1})$
constraint for asymmetric 1-tori. It is also not as strong if the
platonic-solid-like manifolds of compact hyperbolic topologies are
considered, though the overall size of the manifold should be at least
of order the last scattering surface radius \cite{bpss98}.

The primoridial spectral tilt $n_s$ is surprisingly well determined
using CMB data alone, as the upper right panel of Fig.~\ref{fig:CLth}
suggests.  If $\sigma_8$ is marginalized for the tilted $\Lambda$CDM
sequence with $H_0$=50, with DMR only the primordial index is $n_s$ =
$1.02^{+.23}_{-.25}$ with no gravity waves, $\nu_t$=0, and
$1.02^{+.23}_{-.18}$ with gravity waves and $\nu_t$=$\nu_s$, rather
encouraging for the nearly scale invariant models preferred by
inflation theory.  For the 13 Gyr tilted $\Lambda$CDM sequence, when
all of the current CMB data are used \cite{bj98} get
$1.02^{+.05}_{-.03}$ for $H_0=50$ (and $\Omega_{\Lambda}=0$, the
tilted sCDM model sequence) and $1.00^{+.04}_{-.04}$ for $H_0=70$ (and
$\Omega_{\Lambda}=0.66$). Marginalizing over $H_0$ gives
$1.01^{+.05}_{-.04}$ with gravity waves included, $0.98^{+.08}_{-.06}$
if they are not. The marginalized 13 Gyr tilted oCDM sequence gives
$1.00^{+.05}_{-.05}$.

Constraints on such ``global parameters'' as average curvature, $H_0$
and $\Omega_\Lambda$ from COBE data alone are not very good, and the
situation for the $\Lambda$CDM sequences is not that much improved
with all of the CMB data. After marginalizing over all $n_s$,
\cite{bj98} get $H_0 < 75$ at $1\sigma$, but effectively no constraint
at 2$\sigma$. The strong dependence of the position of the acoustic
peaks on $\Omega_k$ means that the oCDM sequence is better restricted:
$\Omega_{tot}\sim .7$ is preferred; for the 13 Gyr sequence this gives
$H_0 \approx 53$ and for the 11 Gyr sequence $H_0 \approx 65$.

\subsection{Cosmic Parameters and LSS (+CMB)} 
\label{secLSSpspec}

We have always combined CMB and LSS data in our quest for viable
models. DMR normalization determines $\sigma_8$ to within 7\%, and
comparing with the $\sigma_8 \sim 0.6\Omega_{nr}^{-0.56} $ target
value derived from cluster abundance observations severely constrains
the cosmological parameters defining the models. In
Fig.~\ref{fig:smprobes}, this means the COBE-normalized ${\cal
P}_{\Phi}(k)$ must thread the ``eye of the needle'' in the
cluster-band.

Similar constrictions arise from galaxy-galaxy and cluster-cluster
clustering observations: the shape of the linear ${\cal P}_\rho $ must
match the shape reconstructed from the data. The one plotted in the
panels of Fig.~\ref{fig:smprobes} is from \cite{peacock96}.  The
clustering observations are roughly compatible with an allowed range
$0.15 \lta \Gamma + \nu_s/2 \lta 0.3$, where the oft-used $\Gamma =
\Gamma_{eq}\, e^{-(\Omega_B(1+\Omega_{nr}^{-1}(2{\rm h})^{1/2})
-0.06)}$, $\Gamma_{eq} \approx \Omega_{nr} \, {\rm h} \,
[\Omega_{er}/(1.68\Omega_{\gamma})]^{-1/2}$, characterizes the density
transfer function shape. The reason $\Gamma_{eq}$ looms so large is
that it parameterizes the scale of the horizon when the energy density
in nonrelativistic and relativistic matter are the same, $k_{Heq}^{-1}
= 5 \, \Gamma_{eq}^{-1} \hmpc $. The sCDM model with $\Omega_B=0.03$
has $\Gamma \approx 0.5$. The phenomenological $\Gamma /\Gamma_{eq}$
correction factor cannot fully parameterize the effects of increasing
$\Omega_B$ on the power spectrum, which has the ``Sakharov
oscillations'' evident in Fig~\ref{fig:smprobes} in the $\Lambda$BCDM
models.

To get $\Gamma + \nu_s/2$ in the observed range one can: lower ${\rm
h}$, lower $\Omega_{nr}$ ($\Lambda$CDM, oCDM), raise $\Omega_{er}$,
the density parameter in relativistic particles ($1.68\Omega_{\gamma}$
with 3 species of massless neutrinos and the photons), \eg as in
$\tau$CDM, with a decaying $\nu$ of lifetime $\tau_d$ and $\Gamma
\approx 1.08 \Omega_{nr} {\rm h}(1 + 0.96 (m_\nu \tau_d /{\rm keV~
yr})^{2/3})^{-1/2}$; raise $\Omega_B$; tilt $\nu_s < 0$, \eg for sCDM
parameters $0.3 \lta n_s \lta 0.7$ would be required. Adding a hot
dark matter component gives a power spectrum characterized by more
than just $\Gamma$ since the neutrino damping scale enters as well as
$k_{Heq}$, as is clear from Fig.~\ref{fig:smprobes}. In the post-COBE
era, all of these models that lower $\Gamma + \nu_s/2$ have been under
intense investigation to see which survive as the data improves, if
any.

The deviation in the shape derived using the APM data \cite{peacock96} 
from the simple $\Gamma$ law form may require more baroque models,
such as tilted h$\Lambda$CDM models (see \eg \cite{GS98,bj98}).

Combining LSS and all CMB data gives more powerful discrimination
among the theories. The approach \cite{bj98} used to add LSS
information to the CMB likelihood functions was to design prior
probabilities for $\Gamma +\nu_s /2$ and $\sigma_8 \Omega_{nr}^{0.56}
$, reflecting the current observations, but with flexible and generous
non-Gaussian and asymmetric forms to ensure the priors can encompass
possible systematic problems in the LSS data. For example, their
choice for $\sigma_8 \Omega_{nr}^{0.56} $ was relatively flat over the
0.45 to 0.65 range.

Using all of the current CMB data and the LSS priors, for the 13 Gyr
$\Lambda$CDM sequence with gravity waves included, \cite{bj98} get
$n_s=1.00^{+.05}_{-.03}$ and $H_0=72\pm 3$ ($\Omega_{\Lambda}\approx
0.7$), respectively, when $H_0$ and $n_s$ are marginalized; with no
gravity waves, $0.96^{+.07}_{-.05}$ and $H_0=70\pm 3$ are obtained;
and for an h$\Lambda$CDM sequence, with a fixed ratio
$\Omega_{hdm}/\Omega_{nr}=0.2$ for two degenerate massive neutrino
species, $n_s\approx 0.97^{+.02}_{-.02}$ and $H_0 \approx
57^{+5}_{-3}$ are obtained, revealing a slight preference for
$\Omega_{\Lambda} \sim 0.3$.  A slight increase in age above 13 Gyr
lowers $H_0$ to the perhaps more preferable 65. 

For the 13 Gyr oCDM sequence, best fit CMB-only models have $\sigma_8$
too low compared with the cluster abundance requirements, so the joint
CMB+LSS maximum likelihood is substantially below that for
$\Lambda$CDM, and severely challenged by the data just as it is for
the SNIa data. 

The isocurvature CDM models with tilt $\nu_{is}>0$, \eg
\cite{pjep97,LM97}, can more or less fit the ${\cal P}_\rho$ shape
data in Fig.~\ref{fig:smprobes}, but are certainly challenged by the
current CMB data in Fig.~\ref{fig:CLth}. $\Lambda$BCDM models with an
increasing value of $\Omega_B$ do have problems with the CMB data, but
it can be partly overcome with tilt; the deviations in the ${\cal
P}_\rho$ shape from the observed are rather too pronounced though (\eg
\cite{ehss98}).  Calculations of defect models ({\it e.g.} strings and
textures) give ${\cal C}_\ell$'s that do not have the prominent peak
that the data seem to indicate \cite{pst97,allen97}, as
Fig.~\ref{fig:CLth} shows.  These are only a few of the many examples
in which the CMB+LSS data has already narrowed our attention
enormously in structure formation model space.

\section{Nonlinear Probes}

We have mentioned that the magnitude of the linear $\sigma_\rho(k)$ is
a good monitor of the sort of dynamics at the resolution that the
$k$-scale defines. For $\sigma_\rho(k) \lta 0.1$ we are solidly in the
linear regime. The weakly nonlinear regime between 0.1 and
$\sigma_\rho \sim 0.7$ defines the $k$-band largely responsible for
the cosmic web, while the regime between 0.7 and 2 encompasses most of
the virialized objects.

Over time there have been a number of approximation techniques
proposed to mimic LSS dynamics. They usually involve the weakly
nonlinear regime and were designed to be significantly faster than the
full $N$-body calculations.  However, with the advent of very fast
computing, unless they had some sort of semi-analytic counterpart to
enhance the depth of understanding such techniques are of
ever-diminishing importance. The most venerable and still useful is
the Zeldovich approximation, \ie first order Lagrangian perturbation
theory. Second order Lagrangian perturbation theory is much more
accurate, but also harder to implement.  The Zeldovich approximation
with sticking at caustic formation is called adhesion theory
\cite{kpsm92}, utilizing Burgers equation for the velocity. It
successfully describes the architecture of voids, filaments and
clusters but does not deal with interior dynamics of ``stuck
structure'' and has not been very useful for mass estimates. The
frozen potential and frozen flow approximations are attempts to avoid
full potential calculations, but are instrinsically numerical, and do
not work that well.  See \cite{colesvarun95} for a review.

What has been applied with great success is weakly nonlinear
perturbation theory, both Eulerian and Lagrangian, much of it by our
French hosts (see \eg \cite{bouchet96}). Some of the milestones in
this effort are: accurate prediction of the one-point distribution
function (PDF) for the overdensity $\delta$ at ``tree level'' {\it
c.f.} N-body results \cite{bernardeau94,colombi97}; analytic expressions for the
cumulants $S_n = \avrg{\delta^n}/(\sigma_{\rho,NL}^2)^{n-1}$ to ``tree
order'' $S_n \sim {\cal O}(\sigma_\rho^{2n-2})$ agree well with
$N$-body results \cite{BaughGaztanagaEfstathiou95}; calculation of the
variance $\avrg{\delta^2}=\sigma_{\rho,NL}^2$ to ``one-loop'' order
\cite{lokas96,roman96} and, in a tour de force, $S_3$ \cite{roman97}
and the bispectrum \cite{roman98} to this order (the
$\sigma_{\rho,NL}^2$ order correction to the tree level results).

As one heads into somewhat stronger nonlinearity, $\sigma_\rho \gta
0.7$, analytic methods to deal with object collapse are heavily
utilized. The hugely popular, trivial-to-implement, Press-Schechter
(1976) method, as modified by \cite{bcek} to the excursion set theory:
at each point in space, $\delta_L (k) $ is allowed to random walk as
the resolution $\sigma_\rho$ increases, \ie as the scale $k^{-1}$
decreases. Once the $\delta_L (k) $ reaches a redshift-dependent
threshold $f_c(z) $, the ``absorbing barrier'', that point is said to
have collapsed at that redshift and is assigned to a halo with mass
$M$ corresponding to the scale $k$. The flaw \cite{bcek} is evident:
nearby points that belong to the same halo will be assigned to
different mass objects. Thus the amazement, and delight, in the
community that the one-point distribution, $n(M)d\ln M$, so derived
fits that of $N$-body group catalogues so well
\cite{bcek,laceycole,bm96}, even below $M_*$. So do such interesting
constrained mass functions as $n(M_2 , z_2\vert M_1 ,z_1)$ giving the
number density of objects of mass $M_2$ at redshift $z_2$ given that
the region is already within one of mass $M_1$ at redshift $z_1$.

The peak-patch picture \cite{bm96} described below is the natural
generalization of the Press-Schechter method to include non-local
effects, spatial correlations and more natural ways of assigning mass
to halos. It is also the natural generalization of BBKS single-filter
peaks theory \cite{bbks} to allow a mass spectrum and solve the
cloud-in-cloud ({\it i.e.}, peak-in-peak) problem. In it, the
threshold $f_c$ becomes a function of the tidal environment the peak
patch finds itself in. It is also at the heart of our cosmic web
picture.

Another approach to the nonlinear evolution of ${\cal P}_\rho(k)$ is
that of Hamilton \etal \cite{ham91}, where a Lagrangian wavenumber
$k_L$ of linear theory is mapped onto a nonlinear wavenumber $k_{NL}$
through $k_{NL}^{-1} \approx (1+\overline{\delta}_{NL})^{-1/3}
k_L^{-1}$, where $\overline{\delta}_{NL}$ is the average nonlinear
overdensity in a mean field region of scale $k_{NL}^{-1}$, just what
you would expect if you took a spherical Lagrangian radius and
compressed it to an Eulerian radius while conserving the mass. A
mapping from the linear power spectrum ${\cal P}_{\rho L}$ to the
nonlinear is defined by ${\cal P}_{\rho NL}(k_{NL})= {\tt fn} ({\cal
P}_{\rho L}(k_L))$, where ${\tt fn} $ denotes a suitably fit nonlinear
function, which, however, is dependent upon the shape of the linear
power spectrum \cite{jmw95}. The VIRGO consortium tested the {\tt fn}
fit given by \cite{peacockdodds96} with their $N$-body runs, and find
accurate mapping from ${\cal P}_{\rho L}$ to power levels ${\cal
P}_{\rho NL}$ up to $\sim 1000$.  It was used by \cite{peacock96} to
take the nonlinear galaxy power spectrum derived \eg from the APM
survey and make an estimate of the linear one: this is what we adopted
for comparing with theory in Fig.~\ref{fig:smprobes}.

There is of course the direct numerical approach to nonlinear
physics. We describe the various $N$-body and hydro methods currently
employed in cosmology. We then turn our attention to the peak-patch
picture and the cosmic web.

\subsection{N-Body and Gas Simulations}
\label{sec:densescl}

The ITP Cluster Comparison of Cosmological Hydro Codes was a
``homework problem'' assigned to simulators at the extended workshop
on Galaxy Formation and Cosmic Radiation Backgrounds held at the
Institute for Theoretical Physics in Santa Barbara in
1995. Calculations were finished in 1996/97 and the paper was
submitted in 1998 \cite{itp95cl}. A constrained single peak field 
was used for initial conditions to ensure that a massive COMA-like
cluster would form at the centre of the simulation.  The results
provide a good snapshot of the $N$-body and hydro methods currently in
use in cosmology, although improvements in all of the codes have been
made since the test. The people participating and the codes used are
listed in Table~\ref{tab:hydro}.
\begin{table}[!htb]
ITP Cluster Comparison of Cosmological Hydro+N-body Codes \cite{itp95cl}
\small
\begin{tabular}{|lllll|} 
\hline 
Group & Method & CPU & machine & storage  \\
\hline 
Bond \& Wadsley  & SPH+P$^3$MG & 119hr & DEC$\alpha$ & 100MB\\
Bryan \& Norman & PPM+PM & 200 & SGI PowCh & 500 \\
Cen & TVD+PM & 5312 & IBM Sp2 & 4400 \\
Evrard & SPH+P$^3$M & 320 & HP375 & 17 \\
Gnedin & SLH+P$^3$M & 136 & SGI PowCh & 90 \\ 
Jenkins, Thomas \& Pearce & SPH+AP$^3$M & 5000 & Cray-T3D &
512 \\
Owen \& Villumsen & ASPH+PM & 40 & Cray-YMP & 106 \\
Navarro & SPH+Direct & 120 & Sparc10+Grape & 75\\ 
 Pen & MMH+MMPM & 480 & SGI PowCh & 900 \\ 
Steinmetz & SPH+Direct & 28 & Sparc10+Grape & 22\\
Couchman & SPH+AP$^3$M & 77 & DEC$\alpha$ & 95 \\
Yepes \&  Klypin & FCT+PM & 350 & Cray-YMP & 480 \\
Warren \& Zurek & Tree & 15360 & Intel-$\Delta$ & 1000 \\
\hline 
\end{tabular}
\label{tab:hydro}

GRAVITY SOLVERS: Direct refers to the brute force direct sum over all
pairs to get the force. Tree refers to organizing distant particles
into groups using a tree. PM refers to particle-mesh, using an FFT to
compute the gravitational potential. MMPM is PM with a Lagrangian mesh
distorted to move with the flow. P$^3$M refers to particle-particle +
PM, using FFTs to get the large scale force, and direct sum to get the
small range force. AP$^3$M is P$^3$M with adaptive mesh refinement, so
not as much costly P-P is needed.  The preferred method of Bond \&
Wadsley is treeP$^3$M, PM~\cite{couchman} for long-range forces with a
tree-like P-P for the short-range. It is twice faster than the
SPH-P$^3$MG code used in our cluster comparison, with more accurate
forces and half the storage requirement. (P$^3$MG refers to
particle-particle plus a multigrid solution to the Poisson equation to
get the long range force.) Warren and Zurek did DM only, but a very
large massively parallel benchmark calculation.

LAGRANGIAN HYDRO SOLVERS: These methods have resolution following the
flow, hence increasing in high densityregions. SPH is Monaghan's
smooth particle hydrodynamics with spatially variable resolution; ASPH
is aspherical SPH, using anisotropic kernels to treat asymmetric
collapses better; MMH is moving-mesh hydro using a pseudo-Lagrangian
grid-based technique; SLH is grid-based softened Lagrangian hydro. The
Jenkins \etal simulation was a highly parallelized calculation using
the SPH code Hydra; Couchman used it on a serial machine.

EULERIAN GRID-BASED HYDRO SOLVERS: PPM is the piecewise parabolic
method; TVD refers to the Total Variation Diminishing hydro scheme;
FCT refers to the Flux-Corrected-Transport scheme. For these Eulerian
schemes to ultimately be effective in such compressible media as we
deal with in cosmology, AMR (adaptive mesh refinement) must be used,
which the PPM code has (an early single AMR version was used). 
\end{table}

In spite of the huge variation in code types, CPU hours devoted,
memory required, mass resolution and spatial resolution used, there
was surprisingly good agreement among the calculations: excellent in
the densities of gas and dark matter; good in the gas temperature,
entropy, pressure, X-ray luminosity density, and gas to dark matter
ratio; not as good in the highly dynamical earlier phases before
cluster virialization.

The largest N-body simulation to date has been done by the VIRGO
consortium, 1 billion particles using a P$^3$M code of a periodic
region $(2000 \hmpc)^3$ in size, on a CRAY T3E with 512 CPUs, of a
$\Gamma=0.21$, $\sigma_8=0.6$ $\tau$CDM model. The mass resolution
defined by the mass per particle was $4\times 10^{12} \msun$ and the
gravitational softening was $200 \kpc$, chosen to ensure a do-able
calculation by avoiding a large P-P slowdown. The texture of the
cosmic web is very evident in the results~\cite{colberg98}.

We have adopted a different approach we refer to as {\it importance
sampling} \cite{bwKing97}, in which we do a number of simulations of
constrained field initial conditions, chosen to sample most
efficiently the statistical distribution of whatever quantities we
wish to probe. It is as if we do a big-box simulation, and go around
sampling this patch and that until we get statistical convergence.
From this point of view, the big box simulations oversample the
patches with large scale fluctuations near the {\it rms}, yet may
undersample the rare protosupercluster or protosupervoid regions. If
the periodic simulation volumes are not big enough so that there is
negligible density power on the scale of the fundamental mode, $k^{-1}
= L/(2\pi)$, then the big-box results will be in error. This is not a
problem for the VIRGO simulation, but is a major headache for
simulations of the Lyman $\alpha$ forest where $\sigma_\rho(k)$
changes slowly, and even more so for simulations of the ``1st *''
region. We design our simulation patches to avoid this
\cite{bwprep}. For many purposes, we have found it adequate to
calculate ``shearing patches'', constrained by specifying the tidal
tensor $\Phi_{,ij}$ at the origin smoothed on a few scales. However,
we also may be interested in simulating complex regions, where more
controlling constraints are used, \eg associated with many
cluster-scale density peaks. Our example of a dense supercluster is of
this form.

We now describe some aspects of simulation design. We first decide on
the mass resolution we wish to achieve. This is set by the lattice
spacing of particles on the initial high resolution grid, $a_L$,
chosen to be $\sim 2 \mpc$ to ensure that the waves needed to treat
the target objects forming in the simulation will be adequate, in this
case clusters. Next we need to determine the spatial resolution, of
the gas and of the gravitational forces, preferably highly linked. In
Lagrangian codes like we use, this varies considerably, being very
high in cluster cores, moderate in filaments, and not that good in
voids. Here we wanted to get good calculations of the cluster cores,
so we wanted our resolution to be in excess of $\sim 40 \kpc$, and the
value we get is about $30 \kpc$. Given the target resolution, we then
have to determine how large the high resolution part of the simulation
volume is by CPU limitations on the number of particles we can run in
the desired time. This may mean the high resolution volume may distort
considerably during the simulation. To combat this, layers are added
of progressively lower resolution volumes, to ensure accurate large
scale tides/shearing fields operate on the high resolution patch.  For
the calculations shown, the High Resolution region of interest (grid
spacing $a_L$, $50^3$ sphere) sits within a Medium Resolution region
($2\,a_L$, $40^3$), in turn within a Low Resolution region ($2\,a_L$,
$32^3$). The mean external tide of the entire patch is linearly
evolved and applied during the calculation.

Thus our sample simulation has a high resolution 104 Mpc diameter
patch with $50^3/2$ gas and $50^3/2$ dark matter particles (initial
grid spacing of 100 comoving kpc), surrounded by gas and dark
particles with 8 times the mass to 166 Mpc, in turn surrounded by
``tidal'' particles with 64 times the mass to 266 Mpc. Bigger
calculations are easily do-able without resorting to massive
parallelization; \eg a $100^3/2$ gas and $100^3/2$ dark matter particle
calculation with a total of 1.6 million particles including the medium
and low resolution regions takes about two weeks of SGI Origin 2000
single processor time.

The mass resolution limits the high $k$ power of the waves that can be
laid down in the initial conditions (Nyquist frequency, $\pi/a_L$),
but for aperiodic patches there is no constraint at the low $k$ end:
we use the FFT for high $k$, but a power law sampling for medium $k$,
and a log $k$ sampling for low $k$, the latter two done using SlowFT,
\ie a direct sum over optimally-sampled $k$ values, with the shift
from one type of sampling to another determined by which gives the
minimum volume per mode in $k$-space. By contrast, the periodic big
box calculations are limited by the fundamental mode. 

\begin{figure}
\vspace{-1.5in}
\centerline{
\epsfxsize=1.5in\epsfbox{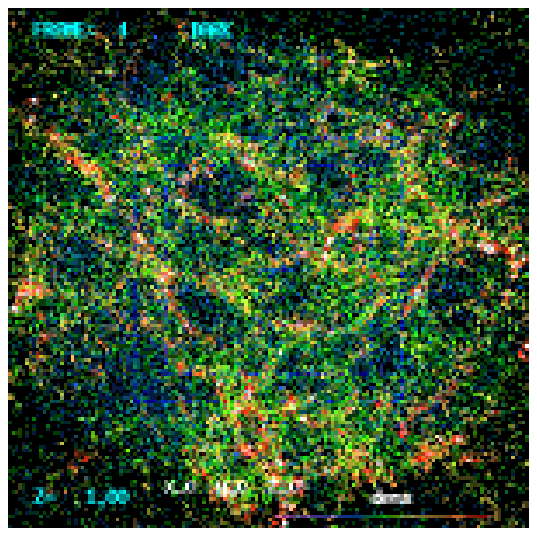}
\hspace{0.0in}
\epsfxsize=1.5in\epsfbox{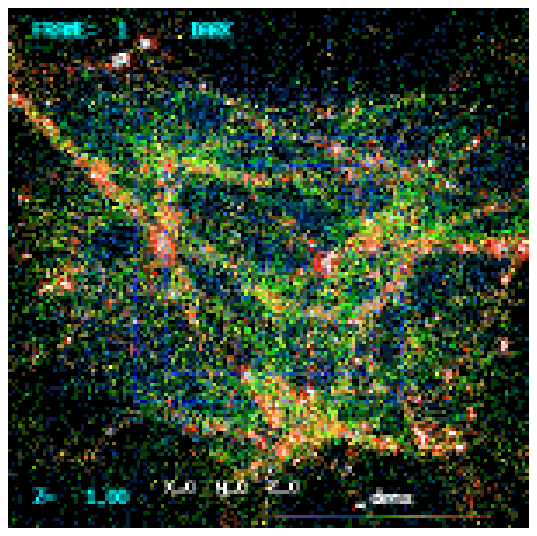}
\hspace{0.0in}
\epsfxsize=1.5in\epsfbox{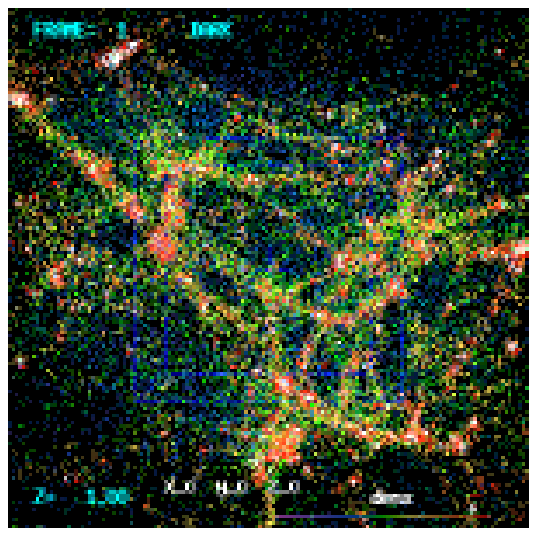}}
\vspace{0.0in}
\centerline{
\epsfxsize=1.5in\epsfbox{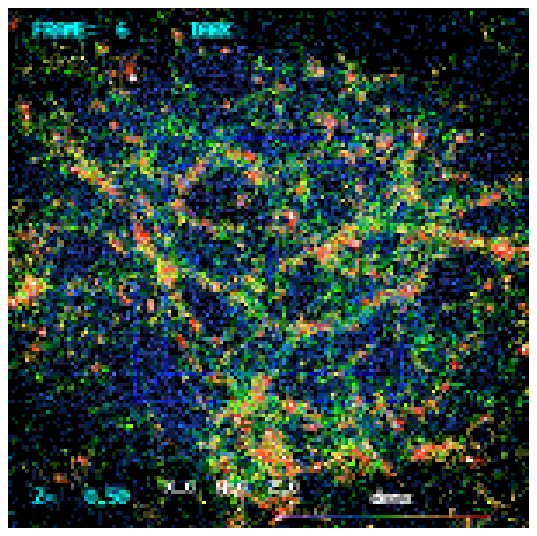}
\hspace{0.0in}
\epsfxsize=1.5in\epsfbox{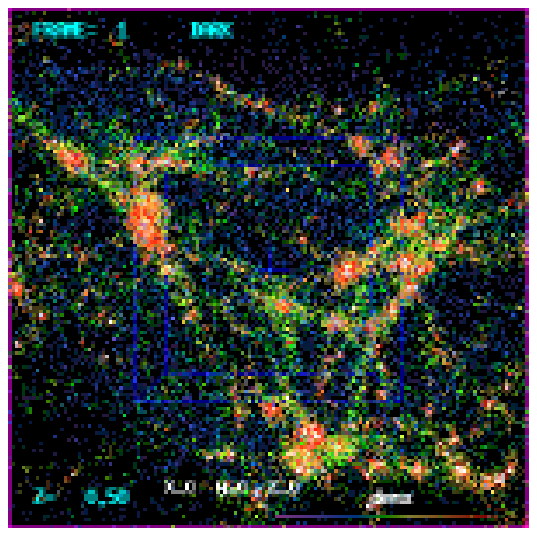}
\hspace{0.0in}
\epsfxsize=1.5in\epsfbox{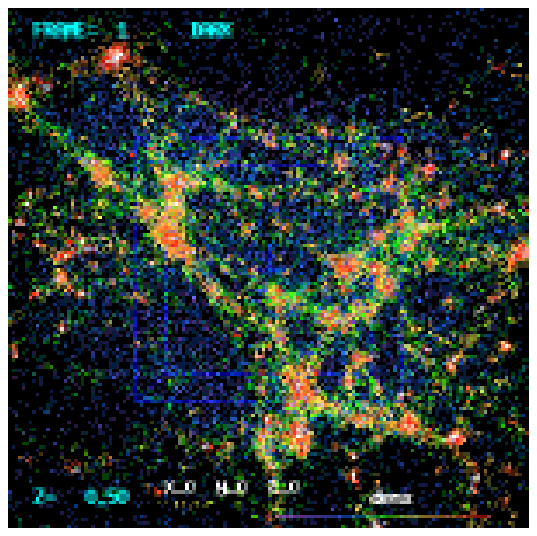}}
\vspace{0.0in}
\centerline{
\epsfxsize=1.5in\epsfbox{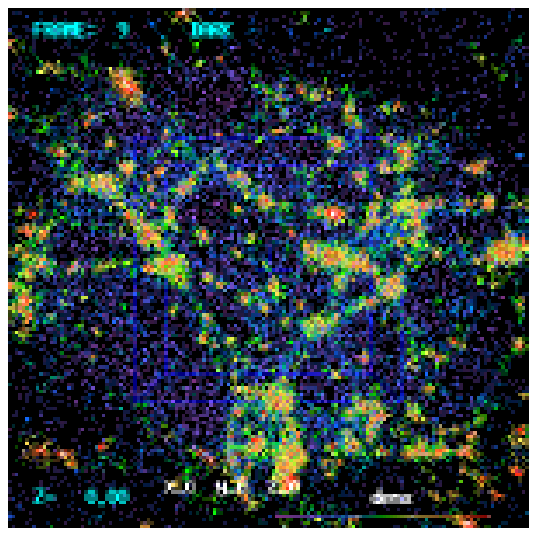}
\hspace{0.0in}
\epsfxsize=1.5in\epsfbox{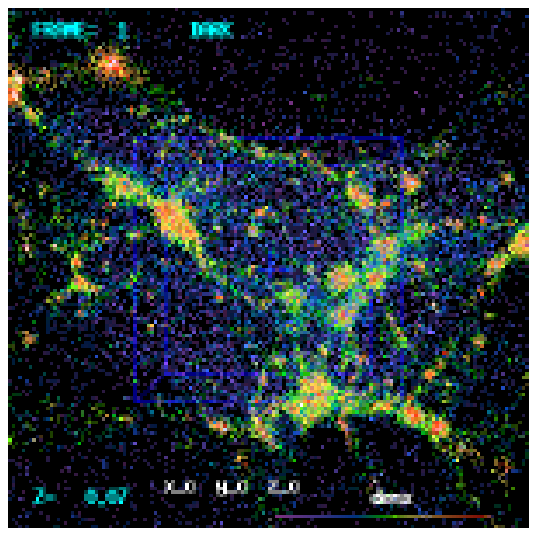}
\hspace{0.0in}
\epsfxsize=1.5in\epsfbox{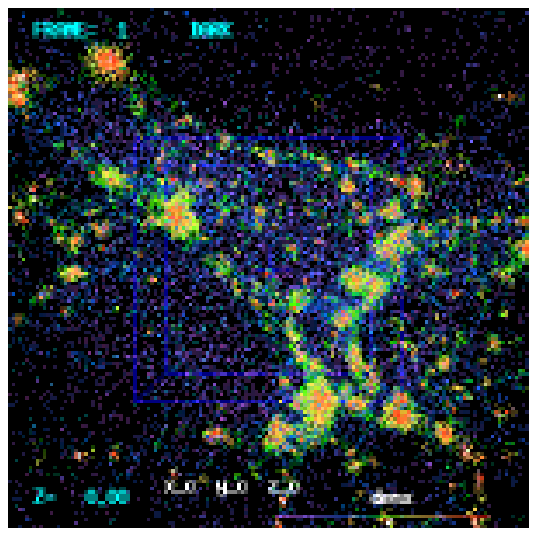}}
\vspace{0.0in}
\caption{COMPACT SUPERCLUSTERS: Dark matter density in a patch $\sim
100 \mpc$ (comoving diameter) across at redshifts $z$=1, 0.5 and 0
(from top to bottom) of three constrained-field supercluster
simulations with differing cosmologies, with HR region $104\mpc$, MR
region $166\mpc$ and LR region $266\mpc$, with wave coverage to
$k^{-1} = 1000\mpc$, the very long waves entering through a self
consistent mean tidal field. The cosmologies shown from left to right
are standard CDM ($\sigma_8$=0.67, ${\rm h}$=0.5), $\Lambda$\'CDM
($\sigma_8$=0.91, ${\rm h}$=0.7, $\Omega_{nr}$=0.3349) and an Open CDM
model ($\sigma_8$=0.91, ${\rm h}$=0.7, $\Omega_{nr}$=0.3689). The
initial conditions were constructed using constraints from the peak
patches either within the region or those exerting a significant tidal
force on it. \eg 17 peak patches (found at $z=0$ in a $400\mpc$
simulation) were used for sCDM. The constraints imply the simulations
have similar, though not identical, patterns. Whereas the middle panel
$\Lambda$CDM simulation had 29, 10 and 2 clusters with mass above
$3\times 10^{14}\msun$ at redshift 0, 0.5 and 1 and oCDM had 20, 12
and 5, sCDM had 13, 3 and none, a dramatic statement of the very
different redshift evolution of the cluster system for sCDM.}
\label{fig:dotplot} 
\end{figure}

The numerical method we adopt is the cosmological SPH+treeP$^3$M code
of Wadsley and Bond \cite{bwprep}. The treeP$^3$M technique is a fast,
flexible method for solving gravity and can accurately treat free
boundary conditions. The code includes photoionization as well as
shock heating and cooling with abundances in chemical (but not
thermal) equilibrium, incorporating all radiative and collisional
processes. The species we consider are H, H$^{+}$, He, He$^{+}$,
He$^{++}$ and e$^{-}$. The simulations were run from $z=30$ to
$z=0$. Parameters for the sCDM, $\Lambda$CDM and oCDM cosmologies we
have run are given in Fig.~\ref{fig:dotplot}.

\begin{figure}
\centerline{
\epsfxsize=5.0in\epsfbox{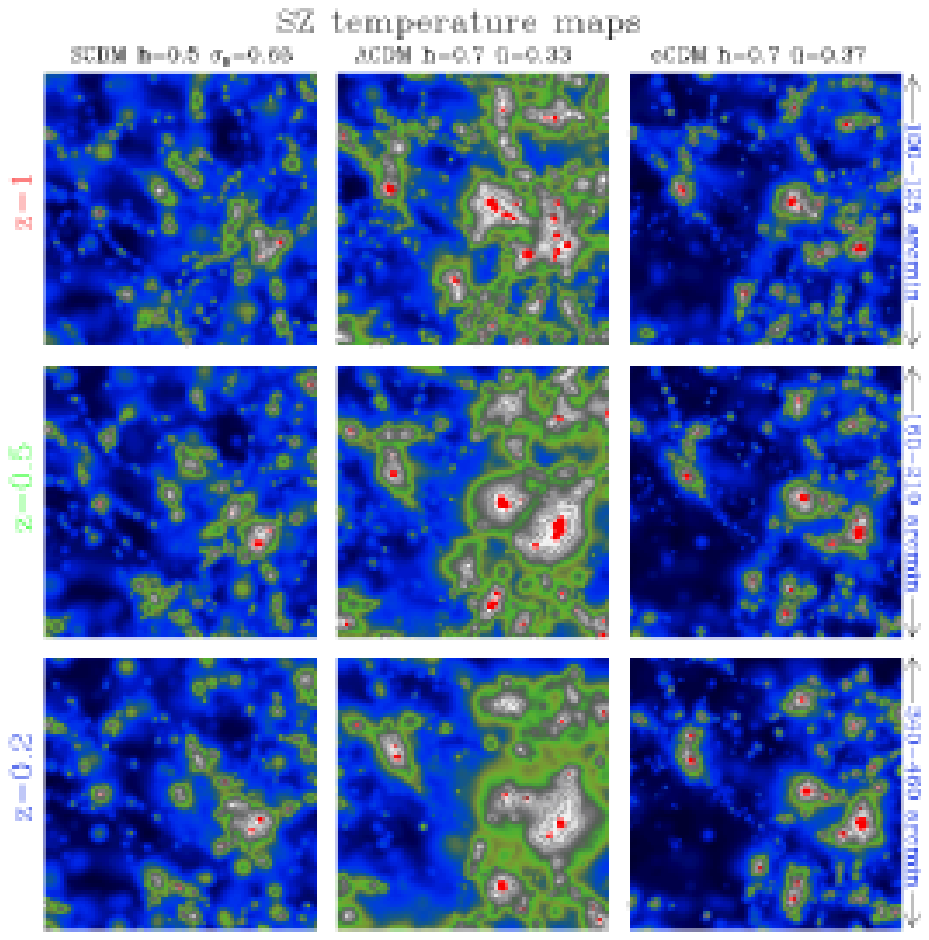}
}
\vspace{0.0in}
\caption{SUNYAEV-ZELDOVICH MAP for the supercluster region seen at
redshifts 1, 0.5 and 0.2 subtending the angles shown. The observing
wavelength was taken to be in the Rayleigh-Jeans region of the
spectrum (so $\Delta T/T =-2y$ here). The core red regions 
are above $32\times 10^{-6}$, and the dark
contours surrounding the white are at $2\times 10^{-6}$. These levels
are now accessible to ground based instrumentation. Blank field SZ
surveys using interferometers and bolometer experiments promise to
revolutionize our approach to the cluster system, especially at $z\gta
0.5$. Although the contours do go into the far field of the cluster
region because they probe the electron pressure, one needs to go below
$10^{-6}$ to fully detect the filaments.  The effect is more
pronounced in the $\Lambda$CDM and $o$CDM universes than in the sCDM
model.  The analogue of this figure in weak lensing is shown in
{\protect\cite{pbkw98}}.}
\label{fig:SZallplot} 
\end{figure}

Our $a_L$ resolution is 2 times better than the VIRGO simulation,
and our best spatial resolution is 6 times better. On the other hand,
we are simulating only $10^{-5}$ of the VIRGO volume at high res: even
with clever sampling, being computer-cheap does have its drawbacks;
though this is a rare event, about a thousand similarly compact
superclusters would be within the VIRGO volume. We are now running
simulations at twice the resolution to ensure accuracy for X-ray
emission in the filaments, not just in the clusters. Our resolution is
adequate for making predictions for signals from the supercluster
complex in the Sunyaev-Zeldovich effect (Fig.~\ref{fig:SZallplot}) and
gravitational lensing (described in our companion paper
\cite{pbkw98}).

\subsection{The Peak-Patch Picture and the Cosmic Web}

We approach the connected ideas we loosely call Cosmic Web theory
(Bond, Kofman and Pogosyan 1996, BKP) and the Peak-Patch Picture (Bond
and Myers 1996, BM, where detailed references to the pre-1995 work
quoted here are given) through a historical path that includes an
outline of the relevant terminology. In 1965, Lin, Mestel \& Shu
showed that a cold triaxial collapse implied an oblate ``pancake''
would form.  In 1970, Zeldovich developed his famous approximation and
argued that pancakes would be the first structures that would form in
the adiabatic baryon-dominated universes popular at that
time. Generally for a cold medium, there is a full non-linear map:
${\bf x} ({\bf r} , t) \equiv {\bf r}-{\bf s} ({\bf r} , t)$, from
Lagrangian (initial state) space, ${\bf r}$, to Eulerian (final state)
space, ${\bf x}$. The map becomes multivalued as nonlinearity develops
in the medium. It is conceptually useful to split the displacement
field, ${\bf s} = {\bf s}_b + {\bf s}_f$, into a smooth quasilinear
long wavelength piece ${\bf s}_b$ and a residual highly nonlinear
fluctuating field ${\bf s}_f$. As we have discussed, if the {\it rms}
density fluctuations smoothed on scale $R_b$, $\sigma_\rho (R_b)$, are
$< {\cal O}(1/2)$, the ${\bf s}_b$-map is one-to-one (single-stream)
except at the rarest high density spots. In the peak-patch approach,
$R_b$ is adaptive, allowing for dynamically hot regions like
protoclusters to have large smoothing and cool regions like voids to
have small smoothing. If $D(t)$ is the linear growth factor, then
${\bf s}_b =D(t) {\bf s}_b ({\bf r} ,0 )$ describes Lagrangian linear
perturbation theory, {\it i.e.} the Zeldovich approximation. The large
scale peculiar velocity is ${\bf V}_{Pb} = -\bar{a}(t) \dot{{\bf
s}}_b({\bf r}, t)$. What is important for us is the strain field (or
deformation tensor): { \small
\begin{eqnarray}
&& e_{b,ij} ({\bf r})\equiv - {1\over 2} \big({\partial s_{bi} \over
\partial r_j}+ {\partial s_{bj} \over \partial r_i}\big) ({\bf r} ) =
-\sum_{A=1}^3 \lambda_{vA} {\hat n}_{vA}^{i} {\hat n}_{vA}^{j},{\rm~where}\nonumber 
\label{eq:eb}\\ && \
\lambda_{v3}={\delta_{Lb}\over 3}(1+3 e_v + p_v),\
\lambda_{v2}={\delta_{Lb}\over 3}(1-2 p_v),\
\lambda_{v1}={\delta_{Lb}\over 3}(1+3 e_v - p_v), \nonumber 
\end{eqnarray}}and $\delta_{Lb}=-e_{b,i}^i$ is the smoothed linear
overdensity, which we often express in terms of the height relative to
the {\it rms} fluctuation level $\sigma_\rho (R_b)$, $\nu_b \equiv
\delta_{Lb}/\sigma_\rho (R_b)$. The deformation eigenvalues are
ordered according to $\lambda_{v3}\ge \lambda_{v2} \ge \lambda_{v1}$
and ${\hat n}_{vA}$ denote the unit vectors of the principal axes. In
that system, $x_A=r_A (1-\lambda_{vA} ({\bf r} , t))$ describes the
local evolution.\footnote{The overdensity $ (1+\delta_Z)({\bf r} ,t
)$= $\vert {(1-D(t)\lambda_{v3} )(1-D(t)\lambda_{v2}
)(1-D(t)\lambda_{v1})}\vert^{-1}$ in a Zeldovich map explodes when the
largest eigenvalue $D(t)\lambda_{v3}$ reaches unity (fold caustic
formation); a pancake develops along the surface ${\hat n}_{v3} \cdot
\nabla_{{\bf r}} \lambda_{v3} =0$. In hierarchical models, classic
Zeldovich pancakes are not relevant for structure formation.}

The strain tensor is related to the peculiar linear tidal tensor by
${\partial^2 \Phi \over \partial x^i \partial x^j} = - 4\pi G {\bar
\rho}_{nr}\bar{a}^2\, e_{b,ij}$, where $\Phi$ is the peculiar
gravitational potential, and to the linear shear tensor by the time
derivative ${\dot e}_{b, ij}$. The anisotropic part of the shear
tensor has two independent parameters, the ellipticity $e_v$ (always
positive) and the prolaticity $p_v$.

Doroshkevich (1973) and later Doroshkevich \& Shandarin (1978) were
among the first to apply the statistics of Gaussian random fields to
cosmology, in particular of $\lambda_{vA}$, at random points in the
medium (see also \cite{pbkw98}). Arnold, Shandarin and Zeldovich
(1982) made the important step of applying the catastrophe theory of
caustics to structure formation. This work suggested the following
formation sequence: pancakes first, followed by filaments and then
clusters.  This should be compared to the BKP Web picture formation
sequence: clusters first, followed by filaments and then walls.  BKP
also showed that filaments are really ribbons, walls are webbing
between filaments in cluster complexes, and that walls are not really
classical pancakes.  For the Universe at $z\sim 3$, massive galaxies
play the role of clusters, and for the Universe at $z\sim 5$ more
modest dwarf galaxies take on that role.

The Web story relies heavily upon the theory of Gaussian random fields
as applied to the rare ``events'' in the medium, {\it e.g.}, high
density peaks. Salient steps in this development begin with BBKS
(1986) \cite{bbks}, where the statistics of peaks were applied to
clusters and galaxies, {\it e.g., } the calculation of the peak-peak
correlation function, $\xi_{pk,pk}$.  In a series of papers, Bond
(1986-90) and Bond \& Myers (1990-93) developed the theory so that it
could calculate the mass function, $n(M)d\ln M$.  It was also applied to
the study of how shear affects cluster alignments ({\it e.g., } the
Binggeli effect), and to Ly$\alpha$ clouds, `Great Attractors', giant
`cluster-patches', galaxy, group and cluster distributions, dusty PGs,
CMB maps and quasars. This culminated in the BM ``Peak-Patch Picture
of Cosmic Catalogues''.

We briefly describe the BM peak-patch method and how it is applied to
initial conditions for simulations such as in
Fig.~\ref{fig:dotplot}. We identify candidate peak points using a
hierarchy of smoothing operations on the linear density field
$\delta_{L}$. To determine patch size and mass we use an ellipsoid
model for the internal patch dynamics, which are very sensitive to the
external tidal field. A byproduct is the internal (binding) energy of
the patch and the orientation of the principal axes of the tidal
tensor. We apply an exclusion algorithm to prevent peak-patch
overlap. For the external dynamics of the patch, we use a
Zeldovich-map with a locally adaptive filter ($R_{pk}$) to find the
velocity ${\bf V}_{pk}$ (with quadratic perturbation theory
corrections sometimes needed). The peaks are rank-ordered by mass (or
internal energy). Thus, for any given region, we have a list of the
most important peaks.  By using the negative of the density field, we
can also get void-patches.

The peak-patch method allows efficient Monte Carlo constructions of 3D
catalogues; gives very good agreement with $N$-body groups; has an
accurate analytic theory with which to estimate peak properties, ({\it
e.g., } mass and binding energy from mean-profiles, using
$f_c(e_v)$, $\langle{e_v \vert \nu_{pk}\rangle}$); and
handles merging, with high redshift peaks being absorbed into low
redshift ones.

\begin{figure}
\vspace{-0.8in}
\centerline{
\epsfxsize=4.0in\epsfbox{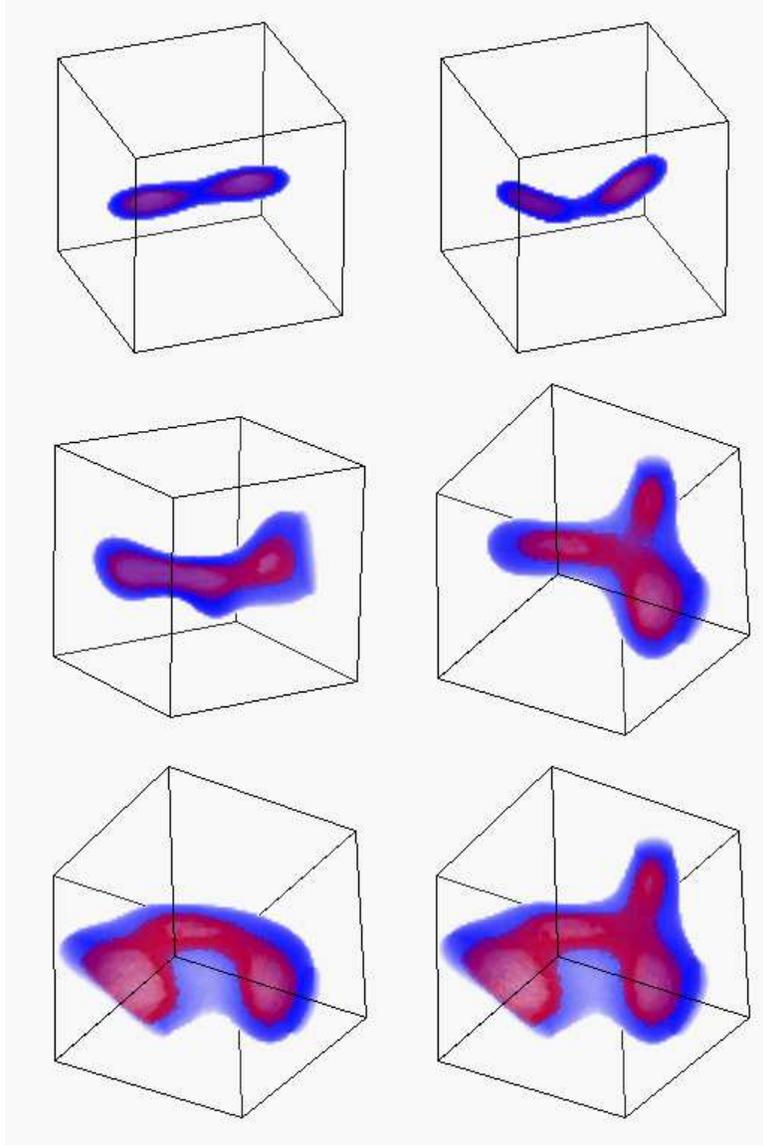}
}
\caption{These plots illustrate the molecular picture of large scale
structure, with ``bonds'' bridging clusters. The initial conditions
have been smoothed and Zeldovich-mapped. (Full $N$-body maps look
similar.) Upper panels show a two-point mean field $\langle \delta_L
\vert 2 $pks$\rangle$ constrained by two oriented clusters separated
by $40\hmpc$, fully aligned and partially aligned. The next 4 panels
show three-point and four-point mean fields for different peak-patch
orientations taken from a simulation. Notice the lower density
contrast webbing between the filaments. One of the supercluster
simulations had initial conditions constrained by the clusters in this
region. }
\label{fig:molecule} 
\end{figure}

BKP concentrated on the impact the peak-patches would have on their
environment and how this can be used to understand the web. They
showed that the final-state filament-dominated web is present in the
initial conditions in the $\delta_{Lb}$ pattern, a pattern largely
determined by the position and primordial tidal fields of rare
events. BKP also showed how 2-point rare-event constraints define
filament sizes (see Fig.~\ref{fig:molecule}). The strongest filaments
are between close peaks whose tidal tensors are nearly aligned, a
binary molecule image with oriented peak-patches as the atoms. Strong
filaments extend only over a few Lagrangian radii of the peaks they
connect. They are so visually impressive in Eulerian space because the
peaks have collapsed by about 5 in radius, leaving the long bridge
between them, whose transverse dimensions have also decreased. The
reason for the strong filaments between aligned peaks is that the high
degree of constructive interference of the density waves required to
make the rare peak-patches, and to preferentially orient them along
the 1-axis, leads to a slower decoherence along the 1-axis than along
the others, and thus a higher density.  Membrane walls are the less
dense webbing between the filaments, a $3,4,\ldots$-peak molecular
image, also shown in Fig.~\ref{fig:molecule}. And void-patches are the
inverse of peak-patches in the initial conditions, but mapped by
nonlinear dynamics into the dominant volume elements.

The Cosmic Web and Peak-Patch pictures provide powerful language for
understanding the structure and evolution of not only cluster systems,
but also galaxy formation and Lyman alpha absorption systems at high
redshift. The web theory predicts the basic structural components of
the IGM as a function of scale, epoch and cosmology. For density
contrasts $\delta \gsim 100$, the rare-events at $z\sim 3$ are massive
galaxies, observed as damped Ly$\alpha$ absorbers, while the typical
collapsed objects are dwarf galaxies responsible for Lyman Limit and
metal line systems. The medium is visually dominated by $\delta \sim
5-10$ filaments, bridging massive galaxy peaks, with smaller scale
filaments within the larger scale ones bridging smaller dwarf galaxies
contributing the most to the $N_{HI} \lta 10^{14.5}$ Ly$\alpha$
forest. Peak-patch galaxy catalogues with all large scale waves
included covering very large volumes of space can be constructed
relatively cheaply for comparison with observations \cite{bwprep}, \eg
with the impressively grand Steidel \etal structures at $z \sim 3.1$
\cite{steidel97}, a task beyond current $N$-body computational
capability. A natural byproduct, of course, is the increasing bias of
halos with increasing velocity dispersion or mass.

We have used field realizations constrained to have interesting
multiple peak/void structures as initial conditions for high
resolution numerical simulations, {\it e.g.} the strong filament of
galaxies in~\cite{bwKing97}. In this paper, we had in mind mimicking
the remarkable Shapley concentration of rich clusters with the
simulations shown in Fig.~\ref{fig:dotplot}. All peak patches tidally
influencing a region that had one of the highest smoothed densities of
protocluster-patches in initial conditions for a $(400 \mpc)^3$
$N$-body simulation were used to set up new initial conditions with
the same basic webbing pattern for our hydro calculations. For the
sCDM case, for example, 17 peaks were used, including the four nearby
clusters shown in the lower panel of Fig.~\ref{fig:molecule} which
exhibited such strong filamentary bridging. These reside in the core
region of Fig.~\ref{fig:dotplot}.

We have descended from ethereal realms of early universe particle
physics, through the calm of linear CMB and matter transport, into the
elegance of the weakly and even strongly nonlinear collisionless
gravitational problems; and finally into gastrophysics, where the
``subgrid physics'' of star formation, multiphase ISM, feedback from
winds, explosions, ionization fronts, \etc must be injected into the
simulatable scales, providing work for decades to come. Even so, with
the amazing databases soon to appear and our developing theoretical
tools, we might just possibly answer how large scale structure formed.

\def\prd{{Phys.~Rev.~D}}
\def\prl{{Phys.~Rev.~Lett.}}
\def\apj{{Ap.~J.}}
\def\apjl{{Ap.~J.~Lett.}}
\def\apjsuppl{{Ap.~J.~Supp.}}
\def\mnras{{M.N.R.A.S.}}

\vfill

\begin{iapbib}{99}{

\bibitem{bh95} Bond,~J.R. 1996, in {\it Cosmology and Large Scale
Structure}, Les Houches Session LX, August 1993, eds. R. Schaeffer
J. Silk, M. Spiro and J. Zinn-Justin (Elsevier Science Press,
Amsterdam), pp. 469-674.

\bibitem{bet97}
Bond,~J.R., Efstathiou,~G.   \& Tegmark,~M.  1997, \mnras\ 291, L33.

\bibitem{zss97}
Zaldarriaga,~M., Spergel,~D. \&  Seljak,~U. 1997, \apj\ 488, 1.

\bibitem{peacock96}
Peacock,~J.A.  1997, \mnras\ 284, 885.

\bibitem{bjk98}
Bond,~J.R., Jaffe,~A.H.  \& Knox,~L. 1998, \apj\ submitted, astro-ph/9808264 

\bibitem{LLKCBA97}
Lidsey,~J.E., Liddle,~A.R., Kolb,~E.W., Copeland,~E.J., Barreiro,~T. \& 
Abney,~M. 1997, Rev. Mod. Phys. 69, 373.

\bibitem{bj98}
Bond, J.R. \&
Jaffe, A. 1998, Proc. Roy. Soc., in press, astro-ph/9809043

\bibitem{huSS97}
Hu,~W., Sugiyama,~N. \& Silk,~J. 1997  Nature\ 386, 37. 

\bibitem{huWhitepol97}
Hu,~W. \& White,~M. 1997, New Astron { 2}, 323.

\bibitem{GS98}
Gawiser, E. \& Silk, J. 1998, preprint. 

\bibitem{bpss98} 
Bond,~J.R., Pogosyan,~D. and Souradeep,~T. 1998, Class. Quant. Grav. {
15}, 2671; see also Cornish,~N.J. \etal, {\it ibid}, 2657;
Levin,~J.J. \etal, {\it ibid}, 2689. 

\bibitem{pjep97}
Peebles, P.J.E. 1997, \apjl\ 483, 1.

\bibitem{LM97}
Linde, A. \& Mukhanov, V.  1997, \prd\ 56, 535.

\bibitem{ehss98}
Eisenstein, D.J., Hu, W., Silk, J. \& Szalay, A. 1998, \apjl\ 494, 1.

\bibitem{pst97}
Pen~Ue-Li, Seljak~U. \& Turok~N. 1997, \prl\ { 79}, 1611.

\bibitem{allen97}
Allen,~B., Caldwell,~R.R., Dodelson,~S., Knox,~L., Shellard,~E.P.S. 
\& Stebbins,~A. 1997, preprint, astro-ph/9704160.

\bibitem{bkp96} Bond, J.R., Kofman, L., \& Pogosyan, D. 1996, Nature
 380, 603.

\bibitem{kpsm92}
Kofman, L., Pogosyan, D., Shandarin, S.F. \& Melott, A.L. 1992, \apj\
393, 437. 

\bibitem{colesvarun95} 
Coles, P. \& Sahni, V. 1995, Phys. Rep. 262, 1.

\bibitem{bouchet96}
Bouchet, F.R. 1996, in Lecture Notes for Course 132 of the E. Fermi
School on Dark Matter in the Universe, astro-ph/9603013

\bibitem{bernardeau94}
Bernardeau, F. 1994, Astron. Ap. 291, 697.

\bibitem{colombi97}
Colombi, S., Bernardeau, F., Bouchet, F.R. \& Hernquist, L.  1997,
\mnras\ 287, 241. 

\bibitem{BaughGaztanagaEfstathiou95}
Baugh, C.M., Gazta\~naga, E. \& Efstathiou, G. 1995, \mnras\ 274, 1049.

\bibitem{lokas96}
Lokas, E.L., Juszkiewicz, R., Bouchet, F.R. \& Hivon, E.  1996, \apj\
467, 1. 

\bibitem{roman96}
Scoccimarro, R. \& Frieman, J.A. 1996, \apj\ 473, 620.

\bibitem{roman97}
Scoccimarro, R. 1997, \apj\ 487, 1. 

\bibitem{roman98}
Scoccimarro, R., Colombi, S., Fry, J.N., Frieman, J.A., Hivon, E. \&
Melott, A. 1998, \apj\ 496, 586. 

\bibitem{ham91}
Hamilton, A.J.S., Kumar, P., Lu, E. \& Matthews, A. 1991, \apjl\ 374,
1. 

\bibitem{jmw95}
Jain, B., Mo, H.J. \& White, S.D.M. 1995, \mnras\ 276, L25. 

\bibitem{peacockdodds96}
Peacock, J.A. \& Dodds, S.J. 1996, \mnras\ 280, L19. 

\bibitem{itp95cl}
Frenk, C.S. \etal 1998, ``The Santa Barbara cluster comparison project:
a test of cosmological hydrodynamics codes'', preprint.

\bibitem{bwKing97} Bond, J.R. \& Wadsley, J.W. 1997, in {\it Computational
   Astrophysics}, p. 323, Proc. 12th Kingston Meeting, ed. D. Clarke \&
   M. West (PASP), astro-ph/970312; Wadsley, J.W. \& Bond, J.R. 1997,
   {\it ibid}, p. 332, astro-ph/9612148; preprint.  

\bibitem{steidel97} 
Steidel, C.C. {\it et al.} 1997, astro-ph/9708125

\bibitem{bm96}
Bond, J.R. \& Myers, S. 1996, \apjsuppl\ 103, 1

\bibitem{bcek}
Bond, J.R., Cole, S., Efstathiou, G. \& Kaiser, N. 1991, \apj\ 379, 440.

\bibitem{laceycole}
Lacey,C.G. \&  Cole, S. 1993, \mnras\ 262, 627; 1994, \mnras\ 271, 676.

\bibitem{bbks}
Bardeen, J., Bond, J.R.,  Kaiser, N. \& Szalay 1986, \apj\ 304, 15.

\bibitem{couchman}
Couchman, H. 1991, \apjl\ 368, 23.

\bibitem{pbkw98}
Pogosyan,~D., Bond,~J.R., Kofman,~L. \& Wadsley,~J.W. 1998, this
volume. 

\bibitem{colberg98}
Colberg, J. \etal (VIRGO consortium) 1998, this volume, astro-ph/9808257;
1997, \mnras\ in press, astro-ph/9711040. 

\bibitem{bwprep} Bond, J.R. \& Wadsley, J.W. 1997, eds Petitjean P. \&
Charlot S., in {\it Proceedings of the XIII IAP Colloquium}, Editions
Frontieres, Paris, p. 143 }
\end{iapbib}
\end{document}